# Structural and Electronic Properties of Graphene and Silicene: An FP-(L)APW+lo Study

Harihar Behera and Gautam Mukhopadhyay

*Department of Physics, Indian Institute of Technology Bombay, Mumbai-400076, India*

**Abstract.** We report here the structural and electronic properties of graphene and silicene (the silicon analogue of graphene) investigated using first-principles calculations of their ground state energies employing full-potential (linearized) augmented plane wave plus local orbital (FP-(L)APW+lo) method. On structure optimization, we found that the graphene-like honeycomb-structure of Si is buckled (buckling parameter $\Delta \simeq 0.44$ Å) in contrast with graphene whose structure is planar ($\Delta = 0.0$ Å). In spite of the buckled-structure, silicene has an electronic structure similar to that of graphene. The results are in agreement with previous reports based on other methods. We have also calculated the lower bounds of the lattice constant "a" of these 2D systems, within the present method of study which are our new results.

## INTRODUCTION

The world of materials research entered an interesting two-dimensional (2D) era [1, 2] with the success in isolating graphene, the two-dimensional monolayer of graphite, from bulk graphite by Novoselov et. al. in 2004 [1]. Graphene, because of its striking properties, has potentials for many novel applications [1–3]. On the other hand, in recent years there have been considerable interest in the fabrication, characterization and properties of silicon nanostructures [4–11]; in particular, silicene, the graphene-like two-dimensional (2D) honeycomb-structure of silicon, has attracted much attention recently both in experiment [5–7] and theory [8–12] for its expected compatibility with contemporary silicon-based micro-electronic industry.

The application of the highly accurate density functional theory (DFT) based full-potential linearized augmented plane wave (FP-LAPW) method [13, 14] in the study of 2D-crystals like graphene is scanty [15]. Therefore, we have carried out an elaborate investigation on the structural and electronic band structure calculations of graphene and silicene using the full-potential (linearized) augmented plane wave plus local orbital (FP(L)APW+lo) method [16, 17], which is a descendant of FP-LAPW method [13, 14], and compared our results with the results from other methods [8–11, 15].

## CALCULATION METHODS

The calculations have been performed by employing the FP-(L)APW+lo method [16,17] within local density approximation(LDA) [18] and Perdew-Burke-Ernzerhof variant of the generalized gradient approximation (GGA-PBE) [19] as implemented in the elk-code [20]. The k-point sampling with $20 \times 20 \times 1$ Monkhorst-Pack (MP) grid was chosen for structural calculation and an MP-grid size of $30 \times 30 \times 1$ was chosen for band-structure calculations. The plane wave cutoff of $|\mathbf{G} + \mathbf{k}|_{max} = 8.5/R_{min}^{MT}$ (a.u.$^{-1}$) for hexagonal carbon (hex-C) and of $|\mathbf{G} + \mathbf{k}|_{max} = 8.5/R_{min}^{MT}$ (a.u.$^{-1}$) for hexagonal silicon (hex-Si) ($R_{min}^{MT}$ is the smallest muffin-tin radius in the unit cell) was chosen for the expansion of the wave functions in the interstitial region, while the charge density was Fourier-expanded up to $|\mathbf{G}|_{max} = 14.5$ (a.u.$^{-1}$). The total energy was converged within 1.36μeV/atom and the ionic relaxation was carried out until the forces acting on atoms decreased bellow 2.5 meV/Å. To simulate the 2D-hexagonal structure of graphene and silicene, 3D-hexagonal supercells with large values of "c" parameter were constructed to keep the interlayer interaction negligibly small. For a fixed value of "c", the ground state energy $E_0$ was calculated for various cell volumes $V$ corresponding to different in-plane lattice constants "a"; then by fitting the $E_0$ vs $V$ data to the Birch-Murnaghan equation of state [21], "a" was extracted from the value of $V$ at the minimum of $E_0$.

## RESULTS AND DISCUSSION

Our results for graphene and silicene along with other relevant information are given in Table 1. As expected, our computed values of "a" for silicene are larger than those of graphene because of larger ionic radius of Si compared to that of C. Also, because of the well-known over binding of the LDA, the GGA-PBE result is larger than the one for LDA. Further, for a graphite monolayer the calculated value of "a" ≈ 2.445 Å is smaller (contracted by 0.6 %) than the experimental value(s) for bulk graphite [22, 23]. It is an experimental fact that the in-plane thermal expansion coefficient of graphite is negative below about 400°C [15, 24]. The explanation [15, 24] that this effect is due to a lateral contraction arising out of the thermal stretching of the crystal along the "c" axis, implies that a monolayer should also exhibit contraction. The in-plane contraction for graphite and hex-Si monolayer are depicted in Fig. 1 and Fig. 2, respectively, which are obtained by calculating "a" by varying "c"; the insets show variation of "a" vs "1/c"; the contracted values of "a" are listed in Table 1. The value of "a" corresponding to c = ∞, obtained by linear fit of the data set (a, 1/c), may be regarded as a lower bound on "a" for the 2D hexagonal structure under study within the present method. Further, our results for graphite monolayer correctly yield the small contraction of "a" as observed experimentally. This contraction is also predicated for a monolayer of hex-Si as shown in Table 1 and Fig. 2. Given this contraction arising out of the value of the "c" parameter used, and basic differences between different other methods, the deviations of our results with others are reasonable.

**TABLE 1.** Calculated ground state results for monolayer of hex-C and hex-Si with lattice constant $|\mathbf{a}| = |\mathbf{b}| = a$. The constant "a", the buckling parameter $\Delta$ and band gap, $E_G$ are compared with reported values. The values "c" (except c = ∞) used for calculation of "a" are given under table-head **c** wherever they are available.

|  | Method | a(Å) | Δ(Å) | $E_G$ (eV) | c (a.u) | Remark |
|---|---|---|---|---|---|---|
| Graphene | LDA | 2.4453 | 0.0 | 0.0 | 20 | This work |
|  | LDA | 2.4449 | 0.0 | 0.0 | 40 | This work |
|  | LDA | 2.4446[*] |  |  | ∞ | This work |
|  | LDA-HGH[†] | 2.4431 | 0.0 | 0.0 | 80 | Pseudopotential [10] |
|  | LDA | 2.450 | 0.0 |  |  | LAPW [15] |
|  | LDA | 2.408 | 0.0 | 0.0 |  | Modified Pseudofunctions [11] |
|  | LDA | 2.46 | 0.0 | 0.0 | 20 | PAW-pot. [9][**] |
|  | GGA-PBE | 2.4576 | 0.0 | 0.0 | 80 | Pseudopotential [10] |
| Silicene | LDA | 3.8111 | 0.4399(9) | 0.0 | 20 | This work |
|  | LDA | 3.8081 | 0.4399(9) | 0.0 | 40 | This work |
|  | LDA | 3.8053 |  |  | ∞ | This work |
|  | LDA | 3.8575 | 0.00 (used)[‡] | 0.0 | 20 | This work |
|  | LDA | 3.8454 | 0.00 (used) | 0.0 | 40 | This work |
|  | LDA | 3.860 | 0.00 (used) | 0.0 |  | PAW-pot. [8] |
|  | LDA | 3.83 | 0.44 | 0.0 | 20 | PAW-pot. [9] |
|  | LDA-HGH | 3.8100 | 0.4247 | 0.0 | 80 | Pseudopotential [10] |
|  | GGA-PBE | 3.8565 | 0.44 (used) | 0.0 | 20 | This work |
|  | GGA-PBE | 3.901 | 0.00 (used) | 0.0 |  | PAW-pot. [8] |
|  | GGA-PBE | 3.8646 | 0.4528 | 0.0 | 80 | Pseudopotential [10] |
| Graphite | Expt. | 2.463 |  |  |  | [22] |
|  | Expt. | 2.456 |  |  |  | [23] |

[*] The value of "a" corresponding to c=∞ is obtained by linear fit of the dataset (a, 1/c) are shown in the inset of Fig.1 for hex-C and Fig.2 for hex-Si. This value of "a" represents the lower bound of "a" for the corresponding 2D hexagonal structure.
[†] Hartwigsen-Goedecker-Hunter variant of Teter-Pade LDA
[**] Projector Augmented Wave Potentials.
[‡] This means that Δ = 0.00 Å is not calculated but used for calculation of the corresponding lattice constant "a".

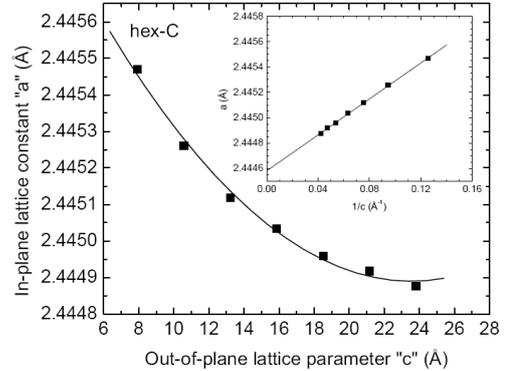

**FIGURE 1:** Calculated lattice constant "a" vs "c" of the hex-C within LDA. Inset shows the "a" vs "1/c".

The band structure plots of 2D hex-C and 2D hex-Si in the buckled structure with a buckling parameter $\Delta \approx$ 0.44 Å, which we obtained from the ionic relaxation within LDA with parameter "c" = 40 a.u., are presented in Fig.3 and Fig.4, respectively (with Fermi energy $E_F$ = 0). In the buckled structure, the locations of the alternating atoms of the hexagonal lattice are in two different parallel planes, i.e., the basal planes in this case;

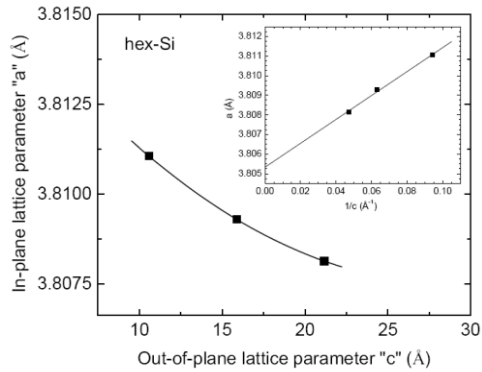

**FIGURE 2:** Calculated lattice constant "a" vs "c" of the hex-Si within LDA. Inset shows the "a" vs "1/c".

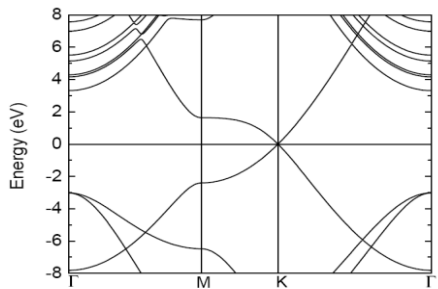

**FIGURE 3:** The LDA result on the band structure of hex-C with c = 40 a.u (graphene); $E_F = 0$.

the buckling parameter $\Delta$ is the perpendicular distance between these two planes. As seen in Fig.4, the energy bands of 2D hex-Si strongly resemble those with the 2D hex-C of Fig.3. In particular, the band gap is zero at the K point of the Brillouin zone and the one-particle energy dispersion around this point is linear. This is the property of the so called Dirac-cone [2, 3] in graphene that is mostly responsible for its unusual exotic property. It is to be noted that this graphene-like electronic structure of 2D hex-Si has recently been predicted theoretically for planar structure (buckling parameter $\Delta = 0.0$ Å) in [8] and for buckled structure ($\Delta \approx 0.44$ Å) in [9, 10], using different methods. Thus the results of our calculations based on a different method corroborate these reported results.

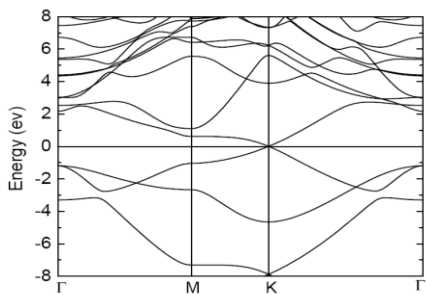

**FIGURE 4:** The LDA result on the band structure of buckled hex-Si with c = 40 a.u. (silicene); $E_F = 0$.

## CONCLUSIONS

We have studied the structural and electronic band structures of monolayers of hex-C and hex-Si (graphene and silicone) by means of DFT based first principles calculations using the FP-(L) APW+lo method. We have found that the buckled hexagonal 2D structure of Si is more stable than the planar one and yet has an electronic structure similar to that of graphene with a linear energy dispersion around the K point. We have also presented lower bounds on "a" for graphene and silicene, which are our new results within the present method of study.